\documentclass[conference]{IEEEtran}
\IEEEoverridecommandlockouts
\usepackage{cite}
\usepackage{amsmath,amssymb,amsfonts}
\usepackage{algorithmic}
\usepackage{graphicx}
\usepackage{textcomp}
\usepackage{xcolor}
\usepackage{array, booktabs, ragged2e}
\usepackage{subcaption}

\usepackage[utf8]{inputenc}
\usepackage[T1]{fontenc}
\usepackage{xparse}

\usepackage{hyperref}
\PassOptionsToPackage{bookmarks=false}{hyperref}
\hypersetup{draft}

\usepackage{footmisc}

\usepackage{listings}
\usepackage{setspace}
\NewDocumentCommand{\codeword}{v}{%
\texttt{\textcolor{black}{#1}}%
}
\usepackage{mathtools}
\usepackage{longtable}
\usepackage{multirow}

\def\BibTeX{{\rm B\kern-.05em{\sc i\kern-.025em b}\kern-.08em
    T\kern-.1667em\lower.7ex\hbox{E}\kern-.125emX}}
\begin{document}

\title{Automatic Lyrics Transcription using Dilated Convolutional Neural Networks with Self-Attention\\
{\footnotesize \textsuperscript{*}}
    \thanks{    
            \protect\includegraphics[clip,width=0.46\textwidth,height=0.88cm]{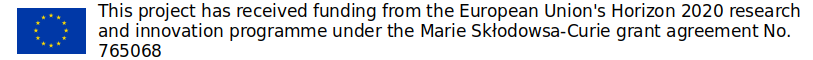}%
}
    }

\author{\IEEEauthorblockN{Emir Demirel}
\IEEEauthorblockA{\textit{Centre for Digital Music} \\
\textit{Queen Mary University of London}\\
London, UK \\
e.demirel@qmul.ac.uk}
\and
\IEEEauthorblockN{Sven Ahlb\"ack}
\IEEEauthorblockA{
\textit{Doremir Music Research AB}\\
\\
Stockholm, Sweden \\
sven.ahlback@doremir.com}
\and
\IEEEauthorblockN{Simon Dixon}
\IEEEauthorblockA{\textit{Centre for Digital Music} \\
\textit{Queen Mary University of London}\\
London, UK \\
s.e.dixon@qmul.ac.uk}
}

\IEEEoverridecommandlockouts
\IEEEpubid{\makebox[\columnwidth]{978-1-7281-6926-2/20/\$31.00~
\copyright2020
IEEE \hfill} \hspace{\columnsep}\makebox[\columnwidth]{ }}

\maketitle

\begin{abstract}

Speech recognition is a well developed research field so that the current state of the art systems are being used in many applications in the software industry, yet as by today, there still does not exist such robust system for the recognition of words and sentences from singing voice.
This paper proposes a complete pipeline for this task which may commonly be referred as automatic lyrics transcription (ALT). We have trained convolutional time-delay neural networks with self-attention on monophonic karaoke recordings using a sequence classification objective for building the acoustic model. The dataset used in this study, \textit{DAMP - Sing! 300x30x2}\cite{b41} is filtered to have songs with only English lyrics. Different language models are tested including MaxEnt and Recurrent Neural Networks based methods which are trained on the lyrics of pop songs in English. An in-depth analysis of the self-attention mechanism is held while tuning its context width and the number of attention heads. Using the best settings, our system achieves notable improvement to the state-of-the-art in ALT and provides a new baseline for the task. 

\end{abstract}

\begin{IEEEkeywords}
automatic speech recognition, machine learning, deep learning, music information retrieval, automatic lyrics transcription, language modeling
\end{IEEEkeywords}

\section{Introduction}
In contemporary pop music, the linguistic content in the singing voice is generally referred as \textit{lyrics} and the process of automatic retrieval this lyrics content from singing voice can then be defined as Automatic Lyrics Transcription\cite{b29}\cite{b33}. The automatic retrieval of pronounced words from speech signals is a widely developed research field and the state of the art systems by today can be successfully applied to industrial applications. However, the same level of robustness has not yet been reached when the input is singing voice. According to prior research, there are several domain specific reasons word recognition performance reduces in singing including domain specific acoustic characteristics \cite{b29,b1} and the alterations of word pronunciations\cite{b7}. Specifically from a machine learning perspective, the main bottleneck for achieving a robust system is the availability of training data with fine-grained annotations, to be used in a supervised learning framework. 

In this study, we exploit such large-scale singing voice dataset, \textit{DAMP - Sing! 300x30x2}\cite{b41} - released by Smule \footnote{Smule is a commercial Karaoke singing application. More info at https://www.smule.com/}, where prompt-level\footnote{In Smule app, lyrics are prompted to the users as words or sentences depending on the song arrangement. Each prompted sentence or word annotation is referred as \textit{prompt-level} annotation.} (and occasionally word-level) annotations are provided. The dataset consists of monophonic Karaoke recordings of pop songs by multiple performers providing near 150 hours of trainable audio data. However, this dataset has not been widely utilized for the purpose of training a word recognition system. Through our proposed framework, we aim to highlight one way of utilization of this dataset in a complete ALT framework and further conduct an in-depth self-attention analysis via fine-tuning experiments.

A robust system for the retrieval of sung lyrics has a variety of potential applications in music information retrieval (MIR) related tasks and the music tech industry. In karaoke and music education apps, the recognition of sung words is essential for tracking a performance and providing feedback to the user. In combination with techniques like query-by-humming, ALT can be utilized for the song identification and metadata retrieval tasks.

Our system uses deep neural networks for building the final acoustic model that is composed of 2D convolutional layers at the front end for extracting more robust features followed by time-delay layers due to their capability of modeling long-term context information. A self-attention layer is added before the final projection layer for weighting the time context when computing the output activations for classification.

Overall, this paper targets at making the following contributions:

\begin{itemize}
    \item Presenting a complete pipeline for ALT in monophonic singing
    \item Testing different language modeling strategies for lyrics
    \item Proposing a neural network architecture combining Self-Attention with CNN and TDNN layers
    \item Performing an in-depth analysis for Self-Attention,
    \item Reporting best results for ALT on a public dataset
    \item Providing the code for open-science and reproducibility

\end{itemize}

This paper is structured as follows: literature on ALT on monophonic singing recordings is reviewed (II). Then the details of the data used in training and evaluation are given (III). Then the proposed system (IV) and the basis of our experiments (V) are explained. The results for each of experimental steps are shown and an in-depth analysis of the self attention parameters is performed (VI). Finally, potential improvements to the proposed system are discussed (VII).

\section{Related Work}

Since the early days of ALT, researchers have shown a tendency to adapting ASR techniques for the retrieval of words or phonemes from the singing voice. Mesaros et al. \cite{b1} adapted a pretrained GMM-HMM based speech recognizer to singing domain using Maximum Linear Likelihood Regression (MLLR) transformation.
In \cite{b3}, the vocal parts were separated, and their proposed method was able to achieve 70\% Phoneme Error Rate (PER) and 88 \% Word Error Rate (WER) on clean and monophonic singing. 

The performance of the ALT systems proposed in recent years has had a notable improvement due to the availability of new open datasets for singing voice. Within the \textit{DAMP} repository\footnote{Can be accessed from \textit{https://ccrma.stanford.edu/damp/}\label{refnote}} there are a few separate open-source datasets for singing made publicly available for research by Smule. The repository consists of Karaoke recordings performed by real-world users of the Smule app. Due to its nature, the audio recordings in the datasets are mostly monophonic accapella singing voice recordings. On one of the earlier releases within the repository, the \textit{DAMP (multiple songs)} \cite{b9} dataset where no line-level lyrics annotations are provided, Kruspe \cite{b29} trained a DNN-HMM system and reported 77\% PER on subset of the aforementioned dataset as the evaluation set. Tsai et. al\cite{b40} use a speech-pretrained TDNN-BiLSTM neural network and retrains on a smaller dataset of 110 monophonic singing recordings obtained from Youtube and reports a WER of 73.09\%. Gupta et al. \cite{b6} adapted a pretrained DNN-HMM based speech recognizer on the \textit{DAMP} dataset and reported 36.32\% WER on a carefully selected subset of this dataset. Further in their work \cite{b7}, they achieved 29.65\% WER by extending the length of pronounced vowels in the pronunciation lexicon. This idea is plausible not only due to the observed performance boost, but also inherently increasing the probability of a frame of a voiced phoneme (i.e phonemes with vowels pronounced) to be followed by another frame with the same phoneme, which is a frequent case in singing.

One of the main challenges of ALT research is the lack of benchmark evaluation sets. Most prior work reports results on different datasets, which makes it harder to compare and evaluate different systems. To overcome this problem, Gerardo et al. \cite{b8} curated a new training and evaluation set, namely the `\textit{DSing Corpus}' based on the \textit{Sing! 300x30x2} dataset\cite{b41} within the \textit{DAMP} repository. On this new evaluation set, the authors reported 19.60\% WER using a factorized Time-Delay Neural Network (TDNN-F) setting and a 4G Language Modeling which is trained on lyrics pop songs in English. To our knowledge, this is the best result reported for ALT from monophonic singing.

\section{Dataset}

All of the experiments of this study are performed on \textit{Sing! 300x30x2} \footref{refnote}. The dataset consists of 18,767 real-world karaoke recordings performed by 13,154 Smule app users, and includes the lyrics prompt timings that are shown to users for the purpose of singing along with the original versions of songs. The karaoke performances in the dataset are unique interpretations of the 300 most popular songs from 30 countries, where the popularity is determined by user votes via the Smule app.

Even though the dataset is curated in a balanced manner and time-level annotations are provided alongside the audio recordings, the data still needs preprocessing to be used for training and evaluation. For instance, the prompt timings are not always in the same granularity, i.e there are both word and sentence level annotations.  Dabike et al. \cite{b8} curated a cleaner version of this dataset based on heuristics, with the goal of removing noisy data. The dataset is also filtered out to keep recordings with lyrics only in English providing nearly 150 hours training data. 

From these recordings, 70 of them with 480 utterances from 43 singers in total are chosen for the \textit{Test} set and 66 songs with 482 utterances from 40 singers for the validation (\textit{Dev}) set, where these recordings from both sets are originating from the UK. In order to meet `the gold standard' for evaluation, the annotations in the \textit{Dev} and \textit{Test} sets are corrected and validated by human experts.

\section{System Details}

Our workflow is based on Kaldi\cite{b42} which is a WFST-based ASR framework that implements the state-of-the-art in research\footnote{Kaldi-ASR is an open-source toolkit and can be accessed from \textit{kaldi\_url}=https://github.com/kaldi-asr/kaldi/blob/master/}. WFST-based speech recognizers  \cite{b37} build a decoding graph composing separate lexicon, acoustic and language models. We use the CMU English Pronunciation Dictionary\footnote{Can be accessed from http://www.speech.cs.cmu.edu/cgi-bin/cmudict} as the lexicon, and train the acoustic and language models via separate neural networks using in-domain data. In this section, we give the details of different components in our automatic lyrics transcription pipeline.

\subsection{Dilated Convolutional Neural Networks with Self-Attention}

For the acoustic model, we propose a deep convolutional neural neural network structure that has three major components: six 2-D convolutional layers at the front-end followed by a stack of sub-sampled factorized Time-Delay Neural Network (TDNN-f) \cite{b18}, and finally, a time-restricted self-attention layer\cite{b43}. The TDNN-f layers are essentially equivalent to dilated 1-D convolutions on the time axis. For this reason, we refer our architecture as a `Dilated Convolutional Neural Network'.

The CNN part consists of six convolutional layers with 3x3 filters. Subsampling (i.e MaxPooling) with a factor of 2 is applied after layers CNN\_3, CNN\_5, CNN\_6, to extract more robust features for TDNN-F layers and to reduce the size of the feature vectors. The rest of the CNN parameters are shown in Table 1.

\begin{table}[!h]
\vskip 0.15in
\label{results}
\setlength\tabcolsep{1.5pt}
\centering
\begin{tabular}{ c |  c  c  c  c  c  c }
\hline
\bfseries  & CNN\_1,2 & CNN\_3 & CNN\_4 & CNN\_5 & CNN\_6  \\ 
\hline
\textit{Height} & 40 & 40 & 20 & 20 & 10 \\
\hline
\textit{Num\_filters} & 48  & 64 & 64 & 64 & 128 \\
\hline
\textit{Subsamp. factor} & N/A &  2 & N/A  &  2  & 2 \\
\hline
\end{tabular}
\smallskip
\caption{Settings of the 2D convolutional layers }
\end{table}

Time Delay Neural Networks (TDNN) are widely used in ASR systems due to their capability of successfully modelling long-term context and makes parallelization possible unlike RNNs. TDNN-F part of the network consists of nine hidden layers with dimension 1024 and a bottleneck dimension of 128. We use the same factorization setting in \cite{b18} for the TDNN-F layers.  Each layer in the network is followed by a ReLU and batch normalization. This architecture without the self-attention layer is set as the baseline NN model in our experiments. We refer this part of the architecture as CTDNN in this paper.

 Since its introduction in the seminal paper \cite{b21}, the self-attention mechanism has been used in many state-of-the-art systems in Natural Language Processing (NLP) \cite{b23}. Recently, this mechanism is also successfully applied in speech recognition\cite{b26} \cite{b27}. Motivated by its success in ASR, we further extend our architecture design by adding a multi-head time-restricted self-attention layer \cite{b43} on top of the network, right before the final linear projection layer. In this attention design, multiple heads learn weights in parallel, focusing on different parts of the time context. Multi-head mechanism is shown to be useful in \cite{b43}. We have set a fixed size for key and query dimensions to 60, value to 40 and tune the context width and the number of heads in the experiments. The models that include the self-attention layer will be referred as CTDNN\_SA.
 
 After the attention layer the weights are projected to one dimensional vectors via \codeword{Linear Layer} continued by softmax. There are two separate output layers that are jointly trained on different objective functions: Maximum Mutual Information (\textbf{Output - chain}) and Cross-Entropy (\textbf{Output - xent}). This joint training is done for regularization as explained in Section V.

\subsection{Language Models}

In our lyrics transcription framework, we use Language Models (LM) that are constructed using in-domain text data that consists of the lyrics of all songs by all artists from UK, USA and Australia in the \textit{DSing30} corpus and the artists from \textit{the Billboard `The Hot 100'} for the years from 2015 to 2018, which is the same training corpus as in \cite{b8}. The training of language models are validated on a set of lyrics from the songs in the validation (\textit{Dev}) set of the DSing30 corpus. The lyrics of the songs in the \textit{Test} set are excluded from the training text corpus when building the LMs. Some statistics of the training text corpus are shown in Table II.

\begin{table}[!h]
\begin{center}
\begin{tabular}{ c c c }
  & Training Corpus & Validation Corpus \\
 \hline
 \textbf{Words} & 7,931,215 & 4018 \\  
 \hline
 \textbf{Sentences} & 1,117,152 & 482 \\
  \hline
 \textbf{Songs} & 4,324 & 66 \\

\end{tabular}
\caption{Some statistics of the LM training \& validation corpora }
\end{center}
\end{table}

In this study, we test three different language models, 3-gram (3G) and 4-gram (4G) LMs train with a maximum entropy objective (MaxENT) built using SRILM \cite{b44} and an LM built using Recurrent Neural Networks (RNNLM) \cite{b45}.

Lyrics may contain made-up words that are not necessary included in the dictionary used in the decoding graph. To handle such words that may exist in the evaluation set, we followed the approach at \footnote{\textit{kaldi\_url}/egs/wsj/s5/utils/lang/make\_unk\_lm.sh}. This procedure helps with the word insertion penalty in scoring when ambigious phone sequences are observed. By ambigious phone sequences, we mean phone sequences that do not exist in the pronunciation dictionary provided. The resulting language model used in decoding will be referred as \textit{n}-G\_\textit{unk} in the experimental results.

\section{Experimental Setup}

We adapt a similar procedure with the `\textit{chain}' recipes\footnote{\textit{kaldi\_url}egs/wsj/s5/local/chain/} in Kaldi for training the neural networks. First a GMM-HMM triphone model is trained to generate phone-level alignments. Based on this model and the alignments, a data cleanup strategy is applied. Using the alignments on the cleaned data, a sequence classifier is trained with the NN architecture proposed in Section IV.

\subsection{GMM-HMM Training}

 For training the GMM-HMM system, we use 13-band MFCCs plus delta and delta-delta features with a frame length of 25 milliseconds and a hop size of 15 milliseconds. $\pm 4$ neighbouring frames are concatenated on the feature vectors (9 consecutive frames in total) to obtain context dependent features. Zero-mean normalization per singer is applied to MFCCs. The spliced frames are then projected to 40 dimensional feature vectors using Linear Discriminant Analysis (LDA). Maximum Likelihood Linear Transformation (MLLT) is applied on the LDA features.  Further, `speaker adaptive training (SAT)' is applied by transforming the feature space (LDA + MLLT) using feature-space maximum likelihood linear regression (fMLLR) per singer\cite{b51}.


A data cleanup strategy is applied with the goal of removing the noise and bad portions of transcripts from the training data using the script at \footnote{\textit{kaldi\_url}/egs/wsj/steps/cleanup/clean\_and\_segment\_data.sh}. We then retrain another GMM-HMM triphone model on the clean training set and generate alignments using this model to be used in neural network training.

\subsection{Neural Network Training}

 The acoustic model is trained using Convolutional Time-Delay Neural Networks (CTDNN) with a lattice-free sequence discriminative training strategy that uses Maximum Mutual Information (MMI) as the objective criterion (LF-MMI) \cite{b14}. A 3-way \textit{speed-perturbation} \cite{b12} is applied for data augmentation, changing the speed of the original signal with the factors of 0.9 and 1.1. Speed perturbation helps to increase the generalizability by modeling different durations for phonemes. Then, we generate lattice alignments as the soft target phone boundaries using the triphone GMM-HMM model prior to training the network. As opposed to GMM-HMM training, we use mel-spectrogram features with 40 filter banks as the input to the neural network. Our framework performs SAT for training the neural networks based on i-vector extraction \cite{b13} as the input speaker (singer) representations with the height of 200. The i-vector speaker representations are projected to smaller matrices with the height of 40 and then combined with filter bank features as separate feature maps but sharing the same trainable weight matrix. The convolutional layers are preceded by a fully connected layer that applies a linear transformation to the input with a trainable matrix to feed to the convolutional layers.

One of the challenges in sequence-level training is that it is prone to overfitting. To overcome this problem we use the combination of three regularization techniques during training \cite{b50}: Cross-entropy regularization, $\textit{l-2}$ regularization and leaky HMMs. In addition, dropout is used at each layer to alleviate the risk of overfitting. Dropout scheduling is used as suggested in \cite{b35}. Training is done on minibatches having sizes of 128 where the data chunks of variable sizes of 140, 100, 160 processed in each minibatch. We use preconditioned Stochastic Gradient Descent (SGD) as the optimizer. The initial and final learning rates are set to $0.0005$ and $0.00005$ respectively, where the learning rate is shrunk after each iteration (minibatch processing). Half of the global learning rate is applied to the last and second-to-last layer weights. We train the network for 8 epochs. After the final iteration of training, we combine models from the last 10 iterations into a final single model using a weighted-average operation \cite{b36}.

\section{Results and Discussion}

In this section, we provide transcription performances of various models studied in this paper. These models include the baseline GMM-HMM model and neural networks with different language models. The effect of number of heads in the attention layer is tested. The training and validation losses of best performing neural network based models are illustrated.

\subsection{GMM-HMM Baseline}

First, we observe how much performance gain can be achieved using a larger dataset by comparing the WERs of GMM-HMM triphone models trained on \textit{DSing1} and \textit{DSing30} corpuses. In Table III, the scores are obtained using the 3G MaxEnt language model. Around 10\% performance gain is observed when using the larger dataset. This improvement is due not only to the larger train set but also because the \textit{DSing30} corpus has more variety in singing accents resulting in learning a more generalizable model.

\begin{table}[!h]
\vskip 0.15in
\label{results}
\setlength\tabcolsep{1.0pt}
\centering
\begin{tabular}{| c | c | c | c |}
\hline
\bfseries Train Set    & \bfseries Dev WER (\%) & \bfseries Test WER (\%) \\
\hline
DSing1 & 63.51  &  63.12 \\
\hline
DSing30 & 52.69 & 50.80 \\

\hline
\end{tabular}
\smallskip
\caption{WERs of Triphone GMM-HMM models on Dev / Test Sets using 3-gram LM. }
\end{table}

\subsection{Language Models}

At this step, our goal is to find the most suitable LM for our task. First, the scoring is done by decoding the FST built using the 3G LM. Then generated lattices are rescored using a graph with 4G LM. Scoring is redone using LMs with `unknown' language modeling. The second-pass scoring is done on RNNLMs for both n-gram models using a pruned rescoring method \cite{b38}. The scores in Table IV are obtained using the baseline CTDNN architecture.

\begin{table}[!h]
\begin{center}

\begin{tabular}{p{1ex}>{\RaggedRight}p{1.5cm}*{3}{p{0.7cm}p{1cm}}}\toprule
\multicolumn{2}{l}{\textbf{LM model}} & \multicolumn{2}{l}{MaxEnt} & \multicolumn{2}{l}{RNNLM} \\\cmidrule{3-4} \cmidrule(lr){5-6} 
 &           & \multicolumn{1}{l}{\textit{Dev}} & \multicolumn{1}{l}{\textit{Test}}  & \multicolumn{1}{l}{\textit{Dev}} & \multicolumn{1}{l}{\textit{Test}} \\
  \cmidrule{3-4} \cmidrule(lr){5-6} \cmidrule{7-8}

 &    3G      &  24.84   &  20.84 &  20.93 & 17.44 \\\midrule

 & 3G\_unk & 24.56 & 20.84 & 20.98 & 17.47 \\\midrule
 & 4G & 21.20 & 18.59 &  18.09 & 16.23 \\\midrule
 & 4G\_unk & \textbf{21.08} & \textbf{18.57} & \textbf{17.70} & \textbf{15.65} \\\midrule

\end{tabular}
\end{center}
\caption{WERs (\%) of different language models using the CTDNN model}
\end{table}

It is seen that RNNLM's outperform the MaxEnt LMs for both n-grams having 2-3 \% WER improvement. Composing the n-gram\_unk language model generally helps with the final WER score.

 \begin{figure*}
    \centering
    \begin{subfigure}[t]{0.95\textwidth}
        \centering
        \includegraphics[width=\textwidth,height=3.82cm]{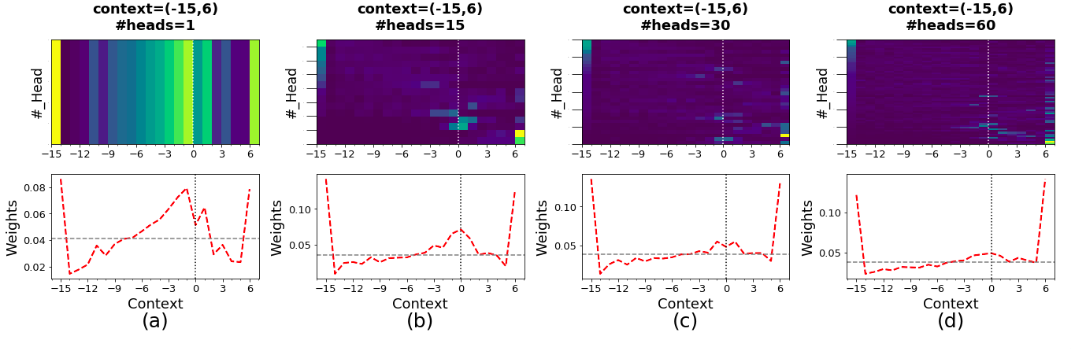}
    \end{subfigure}%
    
    \begin{subfigure}[t]{0.95\textwidth}
        \centering
        \includegraphics[width=\textwidth,height=3.8cm]{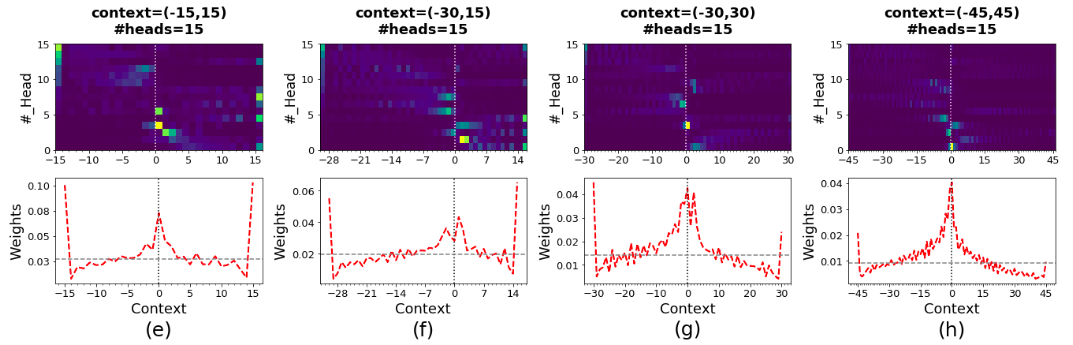}
    \end{subfigure}

    \caption{Illustration of the weight vectors in the self-attention layer. (a,b,c,d) shows the weights for 1, 15, 30, 60 number of heads respectively with a context of (-6,15). Context of (e,f,g,h) are (-15,15), (-30,15), (-30,30), (-45,45) respectively with 15 heads. The figures on top are the the weights for all the heads sorted bottom-to-top w.r.t the weight of leftmost context. The dashed horizontal line in the bottom figures indicates the median of the averaged weights.}
\end{figure*}

\subsection{Neural Networks}

In general, the proposed NN architectures in this study outperform the previously reported score of $23.33$ on \textit{Dev} set and $19.60$ WER (\%) on \textit{Test} set using a 4-gram (4G) MaxEnt language model \cite{b8}. Using the same settings with the aforementioned study, the baseline CTDNN model obtained 18.57\% on the \textit{Test} set showing around 1\% improvement.

The train and validation log-probability losses (vs. iterations) of CTDNN and CTDNN\_SA with different contexts are shown in Figure 2. The losses in models with SA have lower values than the baseline CTDNN model. No explicit sign of overfitting is observed from the train/validation loss plots for any of the models. Early-stopping is not used, yet we have not observed any improvement with further training.

\begin{figure}[!h]
 \centering
 \includegraphics[clip,width=0.47\textwidth,height=2.3cm]{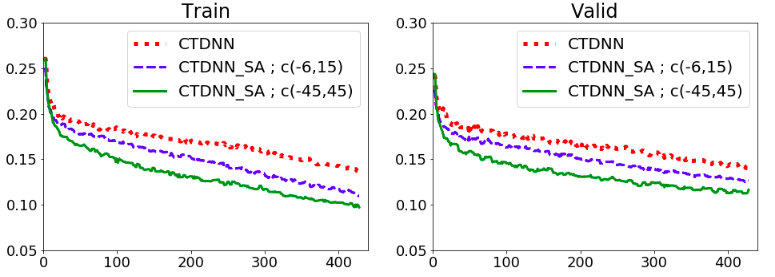}%
 \caption{Log-probability losses of neural network models. `cw' in legends stand for `context'. Both models with SA have 15 heads.'}
 \label{fig:block_diag}
\end{figure}

\subsection{Self-Attention Analysis}

At this stage, we aimed at tuning the self-attention layer by testing the number of heads and the context width by testing  different configurations and performed an in-depth analysis of what the self-attention layer has learned. Rows \# 1 and \#3 in Figure 2 show the attention weight vectors for each head in the self-attention layer and below are the averaged values over all heads.

\begin{table}[!h]
\begin{center}
\begin{tabular}{p{0ex}>{\RaggedRight}p{1.7cm}*{4}{p{1cm}p{0.5cm}}}\toprule
\multicolumn{2}{l}{\textbf{NN Model}} & {\textbf{\#heads}} & \multicolumn{2}{l}{MaxEnt} & \multicolumn{2}{l}{RNNLM} \\\cmidrule{4-5} \cmidrule(lr){6-7} &    &       & \multicolumn{1}{l}{\textit{Dev}} & \multicolumn{1}{l}{\textit{Test}}  &  \multicolumn{1}{l}{\textit{Dev}} & \multicolumn{1}{l}{\textit{Test}} \\
  \cmidrule{4-5} \cmidrule(lr){6-7} \cmidrule{8-9}
 &    TDNN\cite{b8}  & \centering N/A   &  23.33 & 19.60  &  N/A & N/A \\\midrule
 &    CTDNN   & \centering N/A   &   21.08 & 18.57  &  \textbf{17.70} & 15.65 \\\midrule
 & CTDNN\_SA &   \centering 1  & 21.48 & 17.99  & 18.24 & 15.56  \\\midrule
 & CTDNN\_SA & \centering 15 & \textbf{20.38} & \textbf{17.01} &  18.74 & \textbf{14.96} \\\midrule
 & CTDNN\_SA & \centering 30 & 20.46 & 17.79  & 19.49  & 16.13 \\\midrule
  & CTDNN\_SA & \centering 60 & 21.06 & 17.75 & 18.82 & 16.21 \\\midrule

\end{tabular}
\end{center}

\caption{WERs (\%) using different number of heads in the self-attention layer. TDNN-f is given in the table to provide a comparison with the previous best result on the same dataset }

\end{table}

From (a,b,c,d), it is seen that the distribution of the attention weights gets closer to a uniform distribution as the number of heads gets higher, which implies that the self-attention assigns similar weights to context, thus achieving less non-linearity, and reducing the performance. Table V shows the scores of different number of heads in the self-attention layer. These results confirms with the above observation as 15 heads show the best performance. Moreover, the number of trainable parameters increase in the order of millions as the number of heads gets higher (Figure 4).

It is observed that the attention score peaks at the centering bin ($t = 0$) and the weights gradually decrease as getting further from the center. Notice the lower amplitude of the context around the centering bin as the context width gets larger. The attention weights are normalized values due to the softmax layer within the self-attention structure. As the number of context bins increase, the weights assigned to centering bins decrease, while keeping the position of the peaking region in the distribution. The smaller attention weights cause the centering bins have a lesser effect on the final classification. In both of the cases where the context is symmetrical (i.e. the same number of context bins from the past and the future), the decrease of attention scores is slightly sharper for the future bins, implying more attention paid to the past. Finally, Figure 4 shows that the number of trainable parameters increase in the order of 100K for each step we increase the context width. As a conclusion, even though larger context size helps with the training loss (see Figure 3), the WER performance does not improve accordingly as shown in Table VI.

\begin{figure}[!h]
 \centering
 \includegraphics[clip,width=0.4\textwidth,height=3cm]{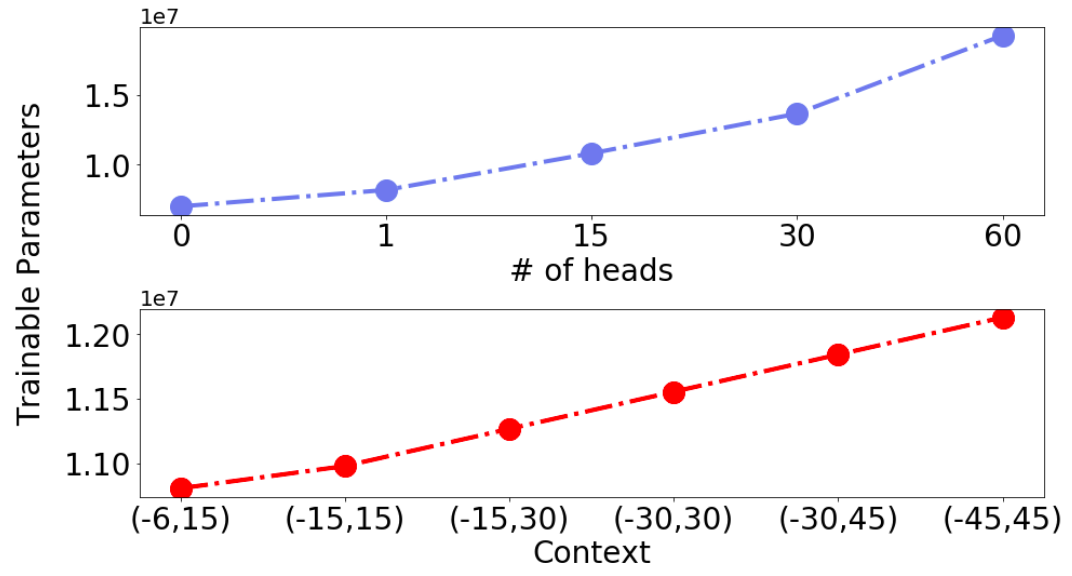}%
 \caption{Number of trainable parameters for each NN model}
 \label{fig:block_diag}
\end{figure}

Overall, setting the context to (-15,6) and number of heads to 15 shows the best WER performance which sets the best score reported in this paper.

\begin{table}[!h]
\vskip 0.15in
\label{results}
\setlength\tabcolsep{1.5pt}
\centering
\begin{tabular}{ c |  c  c  c  c  c  c  }
\hline
\bfseries Context  & (-15,6) & (-15,15) & (-30,15) & (-30,30) & (-45,30) & (-45,45) \\ 
\hline
\textit{Dev} & 18.74 & 18.24 & 18.62 & 18.42 & 18.37 &  \textbf{17.30}  \\
\hline
\textit{Test} & \textbf{14.96}  & 15.61 & 15.93 & 16.26 & 15.13 &  15.26 \\

\hline
\end{tabular}
\smallskip
\caption{CTDNN\_SA scores with different context widths. The results are reported using 4G RNNLM }
\end{table}

\subsection{Decoding Lattices}

Figure 5 illustrates the lattices generated during the decoding stage. The lattices in (a) are obtained using the baseline CTDNN acoustic model with 4G - MaxEnt LM. Adding the self-attention layer to the baseline model results in a much simpler lattice graph (b) reducing the computational complexity and potential confusions during decoding. The graph in (c) is generated by rescoring the lattices using RNNLM where there is only one possible word sequence prediction, boosting the decoding performance even further.

 \begin{figure}[h]
    \centering
    \begin{subfigure}[t]{0.5\textwidth}
        \centering
        \includegraphics[width=\textwidth,height=4.5cm]{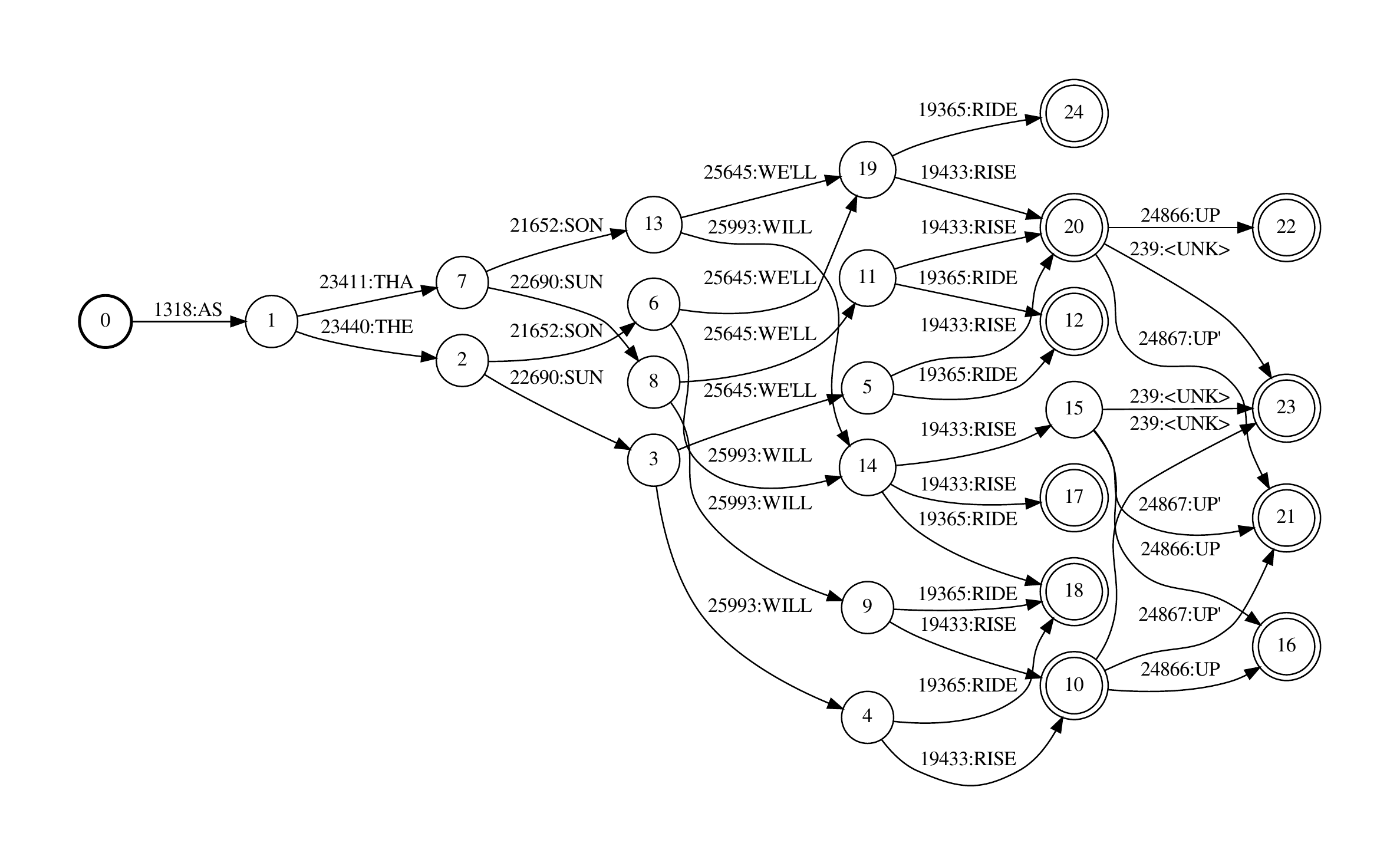}
        \caption{CTDNN - 4G MaxEnt}
    \end{subfigure}%
    
    \begin{subfigure}[t]{0.5\textwidth}
        \centering
        \includegraphics[width=\textwidth,height=2.5cm]{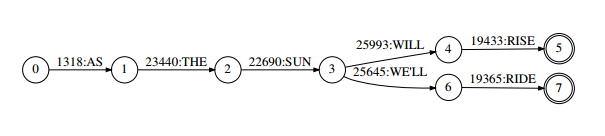}
        \caption{CTDNN\_SA - 4G MaxEnt}
    \end{subfigure}
    
    \begin{subfigure}{0.5\textwidth}
        \centering
        \includegraphics[width=\textwidth,height=2.1cm]{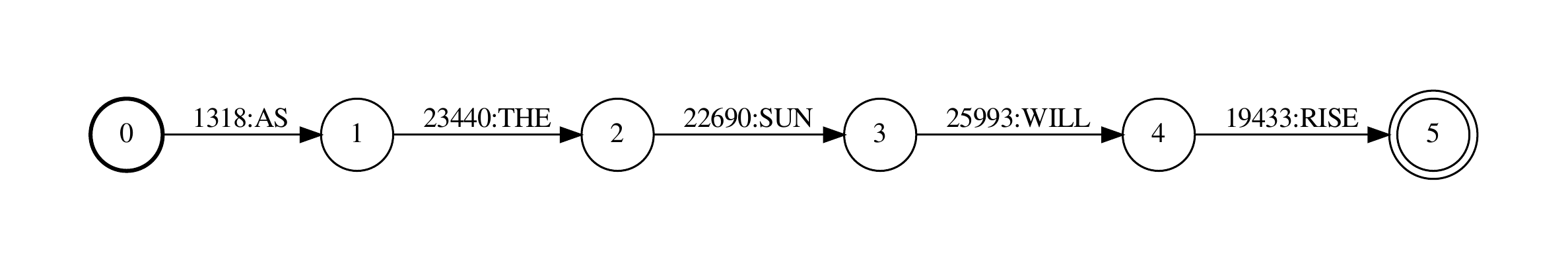}
        \caption{CTDNN\_SA - 4G RNNLM}
    \end{subfigure}
    \caption{Decoding lattices generated with different acoustic and language models for the utterance: `\textit{AS THE SUN WILL RISE}'. The lattice graph in (a) got much simplified in (b) where Self-Attention is used in the neural network. In (c), using RNNLM results in the simplest lattice graph.}.
\end{figure}

Compared to the state-of-the-art in ASR, there is still room for improvement for achieving a robust lyrics transcription system. One of the potential reasons for the high WERs in the experiments might be due to the cleanness of both the training and test data. By inspection, we have observed that the clean-up process described in Section IV does not perfectly clean the dataset from utterances with bad portions or imperfect annotations. In some utterances, the intelligibility of the sung lyrics is not even sufficient for human listeners. For further improvement, the alterations in word pronunciations have to be modified. The lexicon can be modified as suggested in \cite{b7} or learned from data which requires pronunciation annotations in singing recordings. The neural networks can be further tuned and more training data can be acquired through the combination of different datasets.

\section{Conclusion}

Though achieving an approximately 5\% WER improvement to the previously reported best result on the same dataset\cite{b8}, the performance and the generalizability of the proposed system needs further evaluation on different benchmark datasets. Moreover, an evaluation is needed to compare the WER scores of human listeners in order to obtain a performance measure relative to the human level. Considering the current state of the art in ASR, further improvement is necessary to reach a similar robustness level. For reproducibility, we share our code to reproduce the results reported in this paper\footnote{https://github.com/emirdemirel/AutomaticLyricsTranscription-with-Self-Attention}.

The next steps of this research will focus on handling variances in word pronunciations in singing, building a better language model and curating a larger and cleaner training dataset. Unlike spontaneous speech, sung lyrics tend to be pronounced with longer vowels and voiced phonemes. The lexicon used in the decoding graphs needs to be modified in consideration of these alterations in word pronunciations. This new lexicon can be created either be using heuristics\cite{b7} or by learning a pronunciation dictionary from singing data. Language modeling plays an important role in achieving an improved ALT system as shown in our experiments. To achieve this, one might focus on extending/refining the training data and more advanced methods for building the LM. For the former, a larger dataset of lyrics could be constructed and for the latter style-specific LMs can be trained as suggested in \cite{b33}. Finally, the size and variance of the training data for the acoustic model are crucial for performance boost. Training data can be cleaned and new annotations could be generated from weakly or unannotated data using musical heuristics. 

The task of ALT for monophonic singing voice is a task far from being solved, yet recent research indicates a promising future to achieve a robust system. Through this study, we hope to draw more attention to ALT from researchers in both ASR and Music Information Retrieval (MIR) communities, potentially bridging the gap between these two research fields.

\vspace{12pt}

\end{document}